\documentstyle [12pt]{article}
\def\be{\begin{equation}}
\def\ee{\end{equation}}
\def\bea{\begin{eqnarray}}
\def\eea{\end{eqnarray}}
\def\pd{\partial}
\begin{document}

\begin{titlepage}

\title{The Complex Bateman  Equation in a space of arbitrary dimension}

\author{D.B. Fairlie\\
{\it Department of Mathematical Sciences}\\
{\it University of Durham, Durham DH1 3LE}\\
and A.N. Leznov\\
{\it  Institute for High Energy Physics, 142284 Protvino,}\\{\it Moscow Region,
Russia}
{\it and}\\ 
{\it  Bogoliubov Laboratory of Theoretical Physics, JINR,}\\
{\it 141980 Dubna, Moscow Region, Russia}}
\maketitle
\begin{abstract}
A general solution to the Complex Bateman equation in a space of arbitrary 
dimensions is constructed.
\end{abstract}
\end{titlepage}
\section{Introduction}
 We define the Complex Bateman equation in $n$-dimensional space by analogy 
with the so-called Universal Field Equation \cite{fai} as
\be 
\det\left|\begin{array}{cccc}0&\frac{\pd \phi}{\pd\bar y_1}&\ldots&\frac{\pd \phi}
{\pd\bar y_n}\\
\frac{\pd \phi}{\pd y_1}&\frac{\pd^2\phi}{\pd y_1\pd\bar y_1}&\ldots&\frac{\pd^2\phi}{\pd x_1\pd\bar y_n}\\
\vdots&\vdots&\ddots&\vdots\\
\frac{\pd \phi}{\pd y_n}&\frac{\pd^2\phi}{\pd y_n\pd\bar y_1}&\ldots&\frac{\pd^2\phi}{\pd y_n\pd\bar y_n}\\
\end{array}\right|\,=\,0.\label{batman}
\ee
Recently it was shown that the general solution to this equation, in the case where $n=2$
is given implicitly, by equating two arbitrary functions of three variables, $F(\phi,y_1,y_2)$ and $G(\phi,\bar y_1,\bar y_2)$ and solving the resulting equation 
\be
F(\phi,y_1,y_2)\,=\,G(\phi,\bar y_1,\bar y_2).\label{claim} 
\ee
 for $\phi(y_1,y_2,\bar y_1,\bar y_2)$\cite{combat}.  This assertion may be 
 readily verified
by partial differentiation.
The corresponding result for (\ref{batman}) is the subject of this article.

\section{Equivalent First Order Equations}

The complex Bateman equation, (\ref{batman}) is the eliminant of ($n+1$) linear
equations which may be written as:
\be \sum^n_{i=1} \alpha^i \phi_{\bar y_i}=0,\quad \phi_{y_s}=\sum^n_1 
\alpha^i_{y_s} \phi_{\bar y_i}\label{elim1}
\ee
where $\phi_{y_i}$ denotes $\displaystyle{\frac{\pd\phi}{\pd y_i}}$ etc. 
Similarly it is also the eliminant of the equations
\be
\sum^n_{i=1} \beta^i \phi_{y_i}=0,\quad\phi_{\bar y_s}=\sum^n_1 \beta^i_
{\bar y_s} \phi_{y_i}.\label{2}
\ee
From (\ref{elim1}) and (\ref{2}) it follows that:
\be\sum_s \alpha^s_{y_i} \beta^k_{\bar y_s}=\delta^k_{i},\label{3}
\ee
or in other words the Jacobian matrices $\alpha^s_{y_i}$ and  $\beta^k_
{\bar y_s}$ introduced above are inverses of one another. Let us multiply and 
sum each equation of the first system (\ref{elim1}) by $\beta^i$ (and do the same
for  system (\ref{2})). We obtain:
\be
\sum^n_{i=1} \beta^s\alpha^i_{y_s} \phi_{\bar y_i}=0,\quad\sum^n_1 \alpha^s
\beta^i_{\bar y_s} \phi_{y_i}=0\label{4}
\ee

Now we come to a crucial step;  the equations (\ref{4}) cannot contain any
new information, but will only repeat those equations already quoted.
 In other words
\be
\sum_i \beta^i\alpha^s_{y_i}=\theta \alpha^s\label{5}
\ee
for some $\theta$ from which follows immediately  (taking into account the 
inverse properties of the matrices introduced above) in symmetrical fashion,
\be
\sum_i \alpha^i \beta^s_{\bar y_i}=\theta^{-1} \beta^s\label{6}
\ee
which is consistent with the other set of equations. The direct proof of this
proposal will be given in the next section.

Dividing these last equations respectively by $\alpha^s,\beta^s$ and 
introducing the notation $\displaystyle{v^{\nu}=\frac{\alpha^{\nu}}{\alpha_n},
\ u^{\mu}=\frac{\beta^{\mu}}{\beta_n}}$ with the convention that Greek indices 
take values from 1 up to $n-1$
we can eliminate $\theta$ by subtracting the last equation to  arrive at the 
following system:
\be
-v^{\nu}_{y_n}=\sum u^{\mu} v^{\nu}_{y_{\mu}},\quad
-u^{\mu}_{\bar y_n}=\sum v^{\nu} u^{\mu}_{\bar y_{\nu}}.\label{7}
\ee
In the familiar case of two dimensional space this system takes the form
\be
-v_{y_1}=u v_{y_2},\quad -u_{\bar y_1}=v u_{\bar y_2}\label{known}
\ee
the general solution of which is connected with the equation
\be
G(\phi;y_1,y_2)=F(\phi;\bar y_1,\bar y_2)
\ee
We therefore expect that  in the general case of arbitrary
dimensions the general solution of the Complex Bateman equation (\ref{batman})
is connected in some a way with a system of $n-1$ equations for $n-1$ unknown
functions $\psi^\mu$:
\be
Q^{\nu}(\psi_{\mu};y_1,..  y_n)\,=\,P^{\nu}(\psi_{\mu};\bar y_1,..  \bar y_n)
\ \ \nu =1\dots n-1.\label{resultant}
\ee

\section{Conditions of selfconsistency}

As a direct corollary of (\ref{elim1}), (\ref{2}) it follows that the function
$\phi$ can be thought of as either depending upon the set of variables $(u^{\mu};\ y_i)$, or else $(v^{\nu};\ \bar y_j)$
\be
\phi(v^{\nu};\bar y_i)=\phi(u^{\mu};y_j)\label{8}
\ee
This result is equivalent to using  two  equations from (\ref{elim1})
and (\ref{2}). The remaining $2n$ equations, taking into account (\ref{8}) 
may be transformed in the following way:
\bea
\phi_{y_s}&=&\sum \alpha^i_{y_s}\sum \phi_{u^{\mu}}u^{\mu}_{\bar y_i}=
\sum \phi_{u^{\mu}} \sum \alpha^i_{y_s}\left( \frac{\beta^{\mu}_{\bar y_i}}
{\beta^n}-\frac{\beta^{\mu}\beta^n_{\bar y_i}}{ (\beta^n)^2}\right)\nonumber\\
&=&\frac{1}{\beta^n}\sum \phi_{u^{\mu}}(\delta_{s\mu}-u^{\mu}\delta_{sn})\label{inter}
\eea
Or finally:
\bea
\phi_{y_{\mu}}&=&\frac{1}{ \beta^n} \phi_{u^{\mu}}(u^{\mu};y_i),\quad \phi_{y_n}=
-\frac{1}{ \beta^n} \sum u^{\mu} \phi_{u^{\mu}}\label{9}\\
\phi_{\bar y_{\nu}}&=&\frac{1}{ \alpha^n} \phi_{v^{\nu}}(v^{\nu};\bar y_i),\quad \phi_{\bar y_n}=-\frac{1}{\alpha^n} \sum v^{\nu} \phi_{v^{\nu}}\label{10}
\eea
(The reader can compare these equations with the analogous equations in the 
paper on the real Universal equation).
Now let us use the integrability  conditions of selfconsistency  of the  second
order mixed derivatives. Let us assume that conditions of self
consistency for Greek barred and unnbarred indices are satisfied (details
will be given later) :
\be
(\frac{\phi_{v^{\nu}}}{ \alpha^n})_{y_{\mu}}=(\frac{\phi_{u^{\mu}}}{ \beta^n})_
{\bar y_{\nu}}
\ee
and consider what follows from conditions of selfconsistency
for the other pairs $(y_{\mu},\bar y_n),(\bar y_{\nu},y_n)$ and $y_n,\bar y_n$.
We have in consequence, (for the first pair of variables):
$$
\frac{\partial}{\partial y_{\mu}} \frac{\partial \phi}{\partial \bar y_n}=
-\frac{\partial}{\partial y_{\mu}} \sum v^{\nu}\frac{\phi_{v^{\nu}}}
{\alpha^n}=-\frac{\phi_{y_{\mu}}}{ \alpha^n}-\sum v^{\nu} (\frac{\phi_{v^{\nu}}}
{ \alpha^n})_{y_{\mu}}=
$$
\be
-\frac{\phi_{y_{\mu}}}{ \alpha^n}-\sum v^{\nu} (\frac{\phi_{u^{\mu}}}{ \beta^n})
_{\bar y_{\nu}}=\frac{\partial}{\partial \bar y_n}(\frac{\phi_{u^{\mu}}}
{ \beta^n})\label{18}
\ee
The last row of this equality can been transformed into:
\be
\sum_{\alpha} \phi_{u^{\mu},u^{\alpha}}(u^{\alpha}_{\bar y_n}+\sum v^{\nu} 
u^{\alpha}_{\bar y_{\nu}})+(\frac{\phi_{u^{\mu}}}{ \alpha^n})(1-\frac{\sum_i \alpha^i 
\beta^n_{\bar y_i}}{ \beta^n})=0
\ee
Rewriting the equality:
\be
\phi_{\bar y_n}+\sum v^{\nu} \phi_{\bar y_{\nu}}\,=\,0
\ee
in terms of variables  $(u,y)$:
\be
\sum \phi_{u^{\alpha}}(u^{\alpha}_{\bar y_n}+\sum v^{\nu} u^{\alpha}_{\bar 
y_{\nu}})\,=\,0\label{rewrite}
\ee
we arrive at a linear system of $n$ equations for  the $n$ unknowns:
$(u^{\alpha}_{\bar y_n}+\sum v^{\nu} u^{\alpha}_{\bar y_{\nu}}),\quad
{ \alpha^n}(1-\frac{\sum_i \alpha^i \beta^n_{\bar y_i}}{ \beta^n})$,
and assuming that the determinant of the corresponding Universal equation in
the $(n-1)$ dimensional space $u^{\alpha}$ is different from zero
(the degenerate case, when it is equal to zero demands special consideration):
\bea
\sum_i \alpha^i\beta^n_{\bar y_i}&=&\beta^n,\quad 
u^{\mu}_{\bar y_n}+\sum v^{\nu} u^{\mu}_{\bar y_{\nu}}=0\label{24}\\
\sum_i \beta^i\alpha^n_{y_i}&=&\alpha^n,\quad
v^{\nu}_{y_n}+\sum u^{\mu} v^{\nu}_{y_{\mu}}\label{25}
\eea
which proves the previously assumed equations (\ref{6}), (\ref{7}) and shows 
that the hydrodynamical system (\ref{7}) is the direct corollary of the main 
equations. The calculations of the second mixed derivatives $(\phi_{\bar y_n})_
{y_n}$ and in the opposite order leads to equivalent expressions. They are
not essential for what follows and their proof we omit here.

\section{The system of hydrodynamic type }

We understand by a system of hydrodynamic type  the system (\ref{7})  rewritten below :
\be
v^{\nu}_{y_n}+\sum u^{\mu} v^{\nu}_{y_{\mu}}=0,\quad 
u^{\mu}_{\bar y_n}+\sum v^{\nu} u^{\mu}_{\bar y_{\nu}}=0\label{I}
\ee
Two propositions with respect to this system will be required in what
follows.

Proposition 1.

The pair of operators:
\be
D=\frac{\partial}{\partial y_n}+\sum u^{\mu} \frac{\partial}{\partial y_{\mu}},
\quad 
\bar D=\frac{\partial}{\partial \bar y_n}+\sum v^{\nu} \frac{\partial}{\partial 
\bar y_{\nu}}
\ee
are mutually commutative if $(u^{\mu},v^{\nu})$ are solutions of the system 
(\ref{I}).

Acting with the help of operators $(D,\bar D)$ on the second and the first 
equations of (\ref{I}) respectively we come to the conclusion that $2(n-1)$
functions:
\be
\bar D(v^{\nu})=v^{\nu}_{\bar y_n}+\sum v^{\mu} v^{\nu}_{\bar y_{\mu}},\quad 
D(u^{\mu})=u^{\mu}_{y_n}+\sum u^{\nu} u^{\mu}_{y_{\nu}}
\ee
are also solutions of the first and the second system of equations (\ref{I}).

As a corollary we obtain the following  

Proposition 2

\be
v^{\nu}_{\bar y_n}+\sum v^{\mu} v^{\nu}_{y_{\bar \mu}}=Q^{\nu}(v;\bar y),\quad 
u^{\mu}_{y_n}+\sum u^{\nu} u^{\mu}_{y_{\nu}}=P^{\mu}(u;y)\label{BA}
\ee
Indeed the $n$ sets of variables $(1,u)$, and $(1,v)$ respectively satisfy  a 
linear system of 
algebraic equations of $n$ equations, the matrix of which coincides with the 
Jacobian matrix 
\be
J=\det_n \left|\begin{array}{cccc} v^1 & \dots & v^{n-1} & Q^{\nu} \\
                     y_1 & \dots & y_{n-1} & y_n \end{array}\right|
\nonumber
\ee
which in the case of a non-zero solution of the linear system must
vanish. So Proposition 2 is proved.

Compared with (\ref{I}) (\ref{BA}) is  an inhomogeneous system of hydrodynamic
equations separated into functions  $(u,v)$.

Now we are able to find solutions of the primary equations (\ref{24}),(\ref{25}).
To this end, let us rewrite them in the terms of  operators $D_{1,2}$:
\bea
\bar D (\frac{1}{ \beta^n})=-\frac{1}{ \alpha^n \beta^n}=D(\frac{1}{ \alpha^n})
\label{L1}
\eea
We have consequently:
\be
\frac{1}{ \beta^n}=D \Theta,\quad \frac{1}{ \alpha^n}=\bar D\Theta,\quad 
(\bar D\Theta)(D\Theta)=-\bar D D\Theta
\ee 
The solution of the last equation is obvious:
\bea
\exp \Theta=-c(u; y)+\bar c(v; \bar y)\label{L2} 
\eea

\section{General solution of hydrodynamic system}

Let us have the following system of equations defining implicitly $(n-1)$ 
unknown functions $(\psi)$ in $(2n)$ dimensional space $(y,\bar y)$:
\bea 
Q^{\nu}(\psi;y)=P^{\nu}(\psi;\bar y)\label{D}
\eea
with the convention that all Greek indices take values between $1$ and $(n-1)$.
The number of equations in (\ref{D}) coincides with the number of unknown 
functions $\phi^{\alpha}$.

With the help of the usual rules of differentiation of  implicit functions
we find from (\ref{D}):
\be
\psi_y=(P_{\psi}-Q_{\psi})^{-1} Q_y,\quad \psi_{\bar y}=-(P_{\psi}-Q_{\psi})^
{-1}P_{\bar y}\label{DD}
\ee
Let us assume, that between $n$ derivatives with respect to barred and unbarred
variables there exists the linear dependences:
\be
\sum^n_1 c_i \psi^{\alpha}_{y_i}=0,\quad \sum^n_1 d_i \psi^{\alpha}_{\bar y_i}=0
\label{LC}
\ee
and analyse the consequences of these facts.

Assuming that $c_n\neq 0,d_n\neq 0$, dividing  them  into each 
equation of the left and right systems respectively and introducing the
notation $u^{\alpha}=
{c_{\alpha}\over c_n},v^{\alpha}={d_{\alpha}\over d_n}$ we may rewrite the last 
set in the form:
\be
\psi^{\alpha}_{y_n}+\sum^{n-1}_1 u^{\nu} \psi^{\alpha}_{y_{\nu}}=0,\quad
\psi^{\alpha}_{\bar y_n}+\sum^{n-1}_1 v^{\nu} \psi^{\alpha}_{\bar y_{\nu}}=0
\label{MS} 
\ee
Substituting values of the derivatives from (\ref{DD}) and multiplying the
result by the matrix $(P_{\phi}-Q_{\phi})$ from the left we obtain:
\be
Q^{\alpha}_{y_n}+\sum^{n-1}_1 u^{\nu} Q^{\alpha}_{y_{\nu}}=0,\quad
P^{\alpha}_{\bar y_n}+\sum^{n-1}_1 v^{\nu} P^{\alpha}_{\bar y_{\nu}}=0\label{D1}
\ee
From these last equations it immediately follows:
\be
u^{\nu}=-(Q_y)^{-1} Q_{y_n},\quad v^{\nu}=-(P_{\bar y})^{-1} P_{\bar y_n}
\label{UV}
\ee
We see that if we increase the initial system (\ref{D}), by $(n-1)$ vector 
functions
$(u,v)$ defined by (\ref{UV}) then the differential operators  $D,\bar D$
defined by (\ref{3}) in connection with (\ref{MS})  annihilate each 
$\psi$ either as a $Q$ or a $P$ function:
\be
D \psi=\bar D \psi=D Q=D P=\bar D Q=\bar D P=0 \label{VIC}
\ee
This means that $D\bar f(\phi,\bar y)=\bar D f(\phi, y)=0$. And as a direct
corollary of this fact $Dv=\bar D u=0$ and the generators $D,\bar D$ constructed
in this way mutually commute. Thus we have found the general solution of the 
hydrodynamic
system and a concrete realisation of the manifold with the properties of the 
previous section.

With respect to the generators $D,\bar D$ all functions of $2n$ dimensional 
space
may be divided into the following subclasses: functions of general position
$F, D F\neq 0,\bar D F\neq 0$,  holomorphic functions $f,\bar D f=0, D f
\neq 0$, antiholomorphic ones $\bar f, D \bar f=0,\bar d \bar f\neq 0$ and 
$f^0$ "central" functions, both holomorphic and antiholomorphic simultaneously; 
$\bar D f^0=D f^0=0$. Each central function may be represented in the form:
$$
f^0=f^0(Q)=f^0(P)=g^0(\phi)
$$

\section{Equations following from the other restrictions}

Formulae (\ref{UV}) together with (\ref{D}) give the general solution of the
hydrodynamic system (\ref{I}). Indeed this solution depends upon $2(n-1)$ 
arbitrary
functions (\ref{D}) each of $(2n-1)$ independent arguments, which are sufficient
for the statement of Cauchy or Gursat initial value  problems. The general solution of the 
complex 
Bateman equation (\ref{batman})  depends upon only two arbitrary functions 
each of $(2n-1)$ arguments. Thus all other restrictions arising on the way  
must reduce the $2(n-1)$ arbitrary functions of (\ref{D})  to only
two.

For this purpose it is necessary to calculate derivatives of the functions $u,v$ defined by (\ref{UV}). We have in consequence:
\bea
u_{y_{\alpha}}&=&-Q_y^{-1}(Q_{y_n,y_{\alpha}}+\sum Q_{y_n,\phi^{\beta}}\phi^
{\beta}_{y_\alpha}-Q_{y,y_{\alpha}}Q_y^{-1}Q_{y_n}-\sum Q_{y,\phi^{\beta}}\phi^
{\beta}_{y_\alpha}Q_y^{-1}Q_{y_n})\nonumber\\&\equiv&
-Q_y^{-1}(D Q_{y_{\alpha}})+Q_y^{-1}(D Q_{\phi})(P_{\phi}-Q_{\phi})^{-1}
Q_{y_{\alpha}}
\eea
By the same technique we can calculate $u_{\bar y},v_y$ using:
\be
u_{\bar y}=Q_y^{-1}(D Q_{\phi})(P_{\phi}-Q_{\phi})^{-1}P_{\bar y},\quad
v_y=-P_{\bar y}^{-1}(D P_{\phi})(P_{\phi}-Q_{\phi})^{-1}Q_y\label{T}
\ee

Comparing (\ref{8}) with results of the previous section we conclude that the
function $\phi$ is  central  and so depends only on $(n-1)$ arguments
$\phi=\phi(\psi)$. For us it will be more convenient to go back directly to 
linear systems (\ref{elim1}),(\ref{2}) and investigate  their properties. 
 We have in consequence:$$\sum \phi_{\psi^{\alpha}}\psi^{\alpha}_{y_s}=
 \sum \phi_{\psi^{\alpha}}\sum
\psi^{\alpha}_{\bar y_i} \alpha^i_{y_s}
$$
Further evaluation of the last equality is connected with the substitution of
the explicit expressions for the derivatives of the functions $\psi$  (\ref{DD}).
The results of the further calculations we present in the form of multiplication
of the row $\phi_{\psi}$ on the corresponding matrix:
\bea
\phi_{\psi}(P_{\psi}-Q_{\psi})^{-1}(Q_{y_s}+\sum \alpha^i_{y_s} Q_{\bar y_i})=  
\phi_{\psi}(P_{\psi}-Q_{\psi})^{-1}(Q_{y_s}&+&\sum v^{\nu}_{y_s} Q_{\bar y_{\nu}})
\nonumber\\
=\phi_{\psi}(I-\alpha^n(P_{\psi}-Q_{\psi})^{-1}\bar D P_{\psi})(P_{\psi}-
Q_{\psi})^{-1}Q_{y_s}&=&0\label{TTT}
\eea
In the process of the above evaluation  we have used the equalities 
$\bar D Q=0$ and 
the explicit expression for the derivatives of the functions $v$ with 
respect to the unbarred coordinates (\ref{T}).

Equation (\ref{TTT}) for $s=n$ is a direct corollary of the equations with Greek
indices as consequence of the equality $D Q=0$.
Assuming that $\det Q^{\mu}_{y_{\nu}}\neq 0,\ \det (P_{\psi}-Q_{\psi})\neq 0$
we may rewrite the equations (\ref{TTT}) with the Greek indices in the final form:
\be
\sum \phi_{\psi^{\alpha}}(I-\alpha^n(P_{\psi}-Q_{\psi})^{-1}\bar D P_{\psi})^
{\alpha}_{\nu}=0\label{FF}
\ee
A similar equation follows from (\ref{2}): 
\be
\sum \phi_{\psi^{\alpha}}(I-\beta^n(P_{\psi}-Q_{\psi})^{-1}D Q_{\psi})^
{\alpha}_{\nu}=0\label{LF}
\ee

Now we assume that  only one from the set of the $\psi$ functions satisfies
the complex Bateman equation. Suppose it is $\psi^1$ and taht the solution may be chosen in
the form $\phi(\psi^1)$. Of course this suggestion must be confirmed by the 
detailed investigations of all results following from (\ref{FF}) and (\ref{LF}). We omit here this consideration, replacing it by  checking the final 
result. Under this assumption the equations (\ref{FF}) and (\ref{LF}) are 
equivalent to the following $2(n-1)$ equalities: 
$$
(I-\beta^n(P_{\psi}-Q_{\psi})^{-1}D Q_{\psi})^1_{\nu}=0,\quad
(I-\alpha^n(P_{\psi}-Q_{\psi})^{-1}\bar D P_{\psi})^1_{\nu}=0
$$ 
which after substituting into them the explicit expressions for ${1\over \alpha^n}=
\bar D \Theta,{1\over \beta^n}=D \Theta$ from (\ref{L2}) will be convenient 
to rewrite in the form of multiplication of the row on matrix $(P_{\psi}-
Q_{\psi})=\delta$):
$$
(\bar D\Theta,0...0)=(1,0...0)\delta^{-1} \bar D\delta
$$
multiplying the last equality with the matrix $\delta^{-1}$ on the right we 
obtain:
\be
(\bar D\Theta,0...0)\delta^{-1}=-(1,0...0) \bar D\delta^{-1},\quad
(\bar D\Theta,0...0)\delta^{-1}=-(1,0...0) \bar D\delta^{-1}\label{OK}
\ee
(and also a similar system with unbarred differentiation). The integration of 
the last system is straightforward with the result:
\be
(\delta^{-1})_{1,\beta}=\nu(\psi)_{\beta} \exp \Theta={\nu(\psi)_{\beta}\over
(\bar c-c)}\label{OOKK}
\ee

Below we consider the simplest examples of solutions of the last system
for functions $Q,P$  from which the situation in the general case of 
arbitrary $n$ will be clarified.

\subsection{The case $n=2$}

In this case there is only one Greek index and two scalar equations (\ref{OK})
lead to the result:
$$
\exp \Theta=\nu(\psi^1)(P_{\psi^1}-Q_{\psi^1})=\bar c-c
$$
The last equality may be considered as the definition of the functions 
$c,\bar c$  in terms of
$P,Q$:
$$ 
\bar c=\nu(\psi^1)P_{\psi^1}(\psi^1;\bar y_1,\bar y_2),\quad
c=\nu(\psi^1)Q_{\psi^1}(\psi^1,y_1,y_2)
$$
with the correct dependence upon their independent arguments.

\subsection{The case $n=3$}
This case is a more crucial one. Using the explicit form of the matrix $\delta$
 
$$
\delta=\pmatrix{ \Delta^1_{\psi^1} & \Delta^2_{\psi^1} \cr
                 \Delta^1_{\psi^2} & \Delta^1_{\psi^1} \cr}
$$
we without any difficulties can obtain the explicit form of the matrix elements of 
the inverse matrix, which after substitution into (\ref{OOKK}) lead to:
\bea
(P^1_{\psi^2}-Q^1_{\psi^2})+{\nu_2\over \nu_1}(P^2_{\psi^2}-Q^2_{\psi^2})&=&0
\nonumber\\
(P^1_{\psi^1}-Q^1_{\psi^1})+{\nu_2\over \nu_1}(P^2_{\psi^1}-Q^2_{\psi^1})&=&
\bar c-c.
\eea
Taking into account the explicit dependence of $P,Q$ functions upon their
arguments we separate the last system into two systems of equations for $Q,P$
respectively (${\nu_2\over \nu_1}=\nu$):
$$
Q^1_{\psi^1}+\nu Q^2_{\psi^1})=c,\quad Q^1_{\psi^2}+\nu Q^2_{\psi^2})=0
$$
The condition of selfconsistency of the last two equations (equality of
second mixed derivatives of the function $Q^1$ ) leads to
$$
\nu_{\psi^2}Q^2_{\psi^1}-\nu_{\psi^1} Q^2_{\psi^2})=c_{\psi^2}
$$ 
a single equation for the determination of the function $Q^2$ 
Let us consider in the last equation $Q^1$ as an arbitrary given function
$p=p(\psi;y)$. Then  the equation for it may be considered as a definition of
the function $c$, which for what follows it is better to rewrite in the
form:
$$
\nu p_{\psi^1}-(\int d\psi^2 \nu p_{\psi^2})_{\psi^1}=c
$$
Substituting this expression into the first initial equation we obtain for 
$Q^1$:
$$
Q^1=\int d\psi^2 \nu p_{\psi^2}=\nu p-\int d\psi^2 p \nu_{\psi^2}
$$
Introducing the new function $F=\int d\psi^2 p \nu_{\psi^2}$ we now are able 
with its help to represent both functions $Q^1,Q^2$ in a local form:
$$
Q^2={F_{\psi^2}\over \nu_{\psi^2}},\quad Q^1=\nu Q^2-F
$$
The same procedure may to be done with a similar  result with the functions 
$P^1,P^2$:
$$
P^2={\bar F_{\psi^2}\over \nu_{\psi^2}},\quad P^1=\nu P^2-\bar F
$$

Equating $P^{1,2}=Q^{1,2}$ and taking into account that all factors depending 
upon functions $\psi$ may be cancelled we arrive at the following system
\be
F=\bar F,\quad F_{\psi^2}=\bar F_{\psi^2} \label{VFL}
\ee
which determine implicitly two functions $\psi$, one of which, $\psi^1$ is
the solution of the complex Bateman equation in three dimensions. 

\subsection{The general case of arbitrary $n$}

Let us denote the minors of $(n-2)$th order of the first column of the matrix
$\delta, \Delta^1_{\psi^{\beta}}$ by $M^{\beta}$ and introduce the notation
$\nu^{\beta}={M^{\beta}\over M^1}$. Then the systems (\ref{OK}) and (\ref{OOKK})
may be solved in the following way:
\be
\sum \nu^{\alpha} \Delta^{\alpha}_{\psi^1}=\bar c-c,\quad \nu^{\alpha}=
\nu^{\alpha}(\psi)\label{SS3}
\ee
and rewrite the definition of the functions $\nu^{\alpha}$  in the form:
$$
M^{\alpha}=\nu^{\alpha}M^1
$$ 
Multiplying the last equalities by elements of each (given) column
(except for the first one) with further summation of the results we always 
obtain 
zero on the left hand sides of the equalities arising (determinants with
equal columns). Thus we may rewrite (\ref{SS3}) in the equivalent form:
$$
\sum \nu^{\alpha} \Delta^{\alpha}_{\psi^1}=\bar c-c,\quad
\sum \nu^{\alpha} \Delta^{\alpha}_{\psi^A}=0,\quad 2\neq A \neq (n-1)
$$
Keeping in mind that $\Delta^{\alpha}_{\psi^{\beta}}=P^{\alpha}_{\psi^{\beta}}-
Q^{\alpha}_{\psi^{\beta}}$ and recalling the definition of (anti) holomorphic 
functions
of section 5, we separate the last system into two independent ones:
$$
\sum \nu^{\alpha} Q^{\alpha}_{\psi^1}=c,\quad \sum \nu^{\alpha} Q^{\alpha}_
{\psi^A}=0,\quad \sum \nu^{\alpha} P^{\alpha}_{\psi^1}=\bar c,\quad
\sum \nu^{\alpha} P^{\alpha}_{\psi^A}=0
$$
Further transformations of both barred and unbarred systems are similar and so we will follow through the evaluation of the first one.

Introducing new function $F=\sum \nu^{\alpha} Q^{\alpha}$ we obtain:
\be
\sum \nu^{\alpha}_{\psi^1} Q^{\alpha}=-c+F_{\psi^A},\quad 
\sum \nu^{\alpha}_{\psi^A} Q^{\alpha}=F_{\psi^A} \label{SS4}
\ee
In particular we  recall that $\nu^1=1$ and so each equation in its left hand side 
contains only $(n-2)$ terms. Thus between $(n-1)$ equations above at least
one linear dependence exists. Denoting the coefficients of it by $d^A=D^A(\psi)$
we rewrite the condition of self consistency of the last system in the form of
a single equation relating the function $F$ in terms of the function $c$; 
$$
c=F_{\psi^1}+\sum d^A F_{\psi^A}
$$
Inverting the problem we can consider the last equality as a definition
of the function $c$  in terms of the given  $F$. Solving the last $(n-2)$
equations of the system (\ref{SS4}) and adding to them $Q^1$ obtained from
definition of the function $F$, we obtain finally:
$$
Q^B=\sum ( \nu_{\psi}^{-1})^B_A F_{\psi^A},\quad Q^1=F-\sum \nu^A Q^A
$$
Completely similar calculations lead to the following expressions for the 
functions $P$:   
$$
P^B=\sum ( \nu_{\psi}^{-1})^B_A \bar F_{\psi^A},\quad P^1=\bar F-\sum \nu^A P^A
$$ 
We specially emphasize that all coefficient functions in the expressions
above determining  functions $Q,P$ in terms of two arbitrary functions $F,\bar F$
are  functions only of the arguments $\psi$  (central functions) and are the 
same in both cases.

Equating $P^{\beta}=Q^{\beta}$ after obvious cancellation of all (scalar,
matrix) factors depending only upon the functions $\psi$  we reach the final
system of equations implicitly determining all functions $\psi$:
\be
F=\bar F,\quad F_{\psi^A}=\bar F_{\psi^A}\label{FV}
\ee

The function $\psi^1$ is the solution of the Complex Bateman equation 
(\ref{batman}) in the space of $n$ dimension.

\section{The main theorem}

Theorem

THe general solution of the Complex Bateman equation (\ref{batman}) is defined
by the function $\psi^1$, which is implicity determined from the following 
system of $(n-1)$ equations for $(n-1)$ functions $\psi^{\nu}$:
$$
F(\psi;y)=\bar F(\psi;\bar y),\quad F_{\psi^A}=\bar F_{\psi^A},
\quad 2 \leq A\leq (n-1)
$$   

\section{Outlook}

The principal concrete results of the present paper are concentrated in
the Theorem of the previous section, giving the explicit general solution 
(in implicit form) of the complex Bateman equation (\ref{batman}) in $n$
dimensional space.

But by the way a no less important a hydrodynamic type
system with two times $(y_n,\bar y_n)$ was discovered and solved. 
Definite reduction of its general solution
leads to the general solution of Complex Bateman equation. We can't excude the 
possibility that there are exist other reductions leading to no less interesting
systems and equations.

Nevertheless the majority of these results were obtained more on the basis
of intuitive calculations. We think that the algebraic-geometrical form of the
answer tells us of the necessity to consider the group of motion ( the 
symmetry structure) of the manifold (\ref{D}) determining the general solution of 
hydrodynamic system (\ref{9}). The properties of the group of inner symmetry of
it must explain more precisely and directly the proposed way of integration
of the systems under consideration (hydrodynamic and Complex Bateman) as
the uniquely possible one.

We have a feeling that in these problems  algebraic-geometrical methods are 
more appropriate and effective.
Unfortunately they are not within our area of expertise.

\end{document}